\documentclass[reprint,aps,prl,twocolumn,floatfix,secnumarabic,hidelinks]{revtex4-2}
\usepackage{graphicx}
\usepackage{caption}
\usepackage{ragged2e}
\usepackage{svg}
\usepackage{siunitx}
\usepackage{amsmath}
\usepackage{amssymb}
\usepackage[nolist]{acronym}
\usepackage{tabularray}
\UseTblrLibrary{booktabs}
\usepackage{csquotes}
\usepackage{physics}
\usepackage{pifont, xcolor}

\usepackage[hidelinks]{hyperref}

\setcitestyle{super, sort&compress}

\renewcommand{\figurename}{Fig.}
\renewcommand{\tablename}{Table}

\newcommand{\Title}{Distributed Quantum Computing across an Optical Network Link}

\newcommand{\authorcite}[1]{\citeauthor{#1}\cite{#1}}
\newcommand{\ish}{\mbox{$\sim$}\,}

\newcommand{\fig}[1]{Fig.~\ref{#1}}
\newcommand{\extfig}[1]{Ext. Fig.~\ref{#1}}
\newcommand{\tab}[1]{Table~\ref{#1}}
\newcommand{\methods}[1]{\hyperref[#1]{(Methods)}}

\newcommand{\iswap}{iSWAP}
\newcommand{\swap}{SWAP}

\newcommand{\network}{network}
\newcommand{\auxiliary}{auxiliary}
\newcommand{\circuit}{circuit}

\newcommand{\ion}[2]{\mbox{$^{#2}$#1$^+$}}
\newcommand{\Ca}[1]{\ion{Ca}{#1}}
\newcommand{\Sr}[1]{\ion{Sr}{#1}}

\renewcommand{\S}[1]{\text{S}_{#1}}
\renewcommand{\P}[1]{\text{P}_{#1}}
\newcommand{\D}[1]{\text{D}_{#1}}

\newcommand{\down}[1]{0_{#1}}
\newcommand{\up}[1]{1_{#1}}
\newcommand{\Q}[1]{\mathcal{Q}_{#1}}
\newcommand{\net}{\mathrm{N}}
\newcommand{\Qnet}{\Q{\net{}}}
\newcommand{\aux}{\mathrm{X}}
\newcommand{\Qaux}{\Q{\aux{}}}
\newcommand{\cir}{\mathrm{C}}
\newcommand{\Qcir}{\Q{\cir{}}}

\newcommand{\hoa}{HOA2}
\newcommand{\phoenix}{Phoenix}

\newcommand{\qinternet}{\textit{quantum internet}}

\newcommand{\czfidelity}{\SI{86.1(9)}{\percent}} % SPAM-corrected: included in the process tomography POVMs
\newcommand{\iswapfidelity}{\SI{70(2)}{\percent}} % SPAM-corrected: included in the process tomography POVMs
\newcommand{\swapfidelity}{\SI{64(2)}{\percent}} % SPAM-corrected: included in the process tomography POVMs

\newcommand{\groveraverage}{\SI{71(1)}{\percent}}

\newcommand{\rawinfidelity}{\SI{2.85(9)}{\percent}} % Corrected for measurement errors
\newcommand{\rawfidelity}{\SI{97.15(9)}{\percent}} % Raw fidelity to Psi+ Bell-state; corrected for measurement errors
\newcommand{\alicememoryerrors}{\SI{1.9(4)}{\percent}} % SPAM-corrected: included in the process tomography POVMs
\newcommand{\alicememoryfidelity}{\SI{98.1(4)}{\percent}} % SPAM-corrected: included in the process tomography POVMs
\newcommand{\bobmemoryerrors}{\SI{1.8(5)}{\percent}} % SPAM-corrected: included in the process tomography POVMs
\newcommand{\bobmemoryfidelity}{\SI{98.2(5)}{\percent}} % SPAM-corrected: included in the process tomography POVMs

\newcommand{\alicewzzfidelity}{\SI{97.5(2)}{\percent}} % SPAM-corrected
\newcommand{\alicewzz}{\SI{2.5(2)}{\percent}} % SPAM-corrected
\newcommand{\bobwzzfidelity}{\SI{98.0(2)}{\percent}} % SPAM-corrected
\newcommand{\bobwzz}{\SI{2.0(2)}{\percent}} % SPAM-corrected

\newcommand{\alicetransfer}{\SI{0.76(3)}{\percent}} % This is for two transfers
\newcommand{\alicepertransfer}{\num{3.8(2)e-3}} % This is for one transfers
\newcommand{\bobtransfer}{\SI{0.52(1)}{\percent}} % This is for two transfers
\newcommand{\bobpertransfer}{\num{2.6(1)e-3}} % This is for one transfers

\newcommand{\alicesrrotation}{\num{4.8(3)e-4}} % 
\newcommand{\bobsrrotation}{\num{9.8(3)e-4}} % 

\newcommand{\alicemidcircuiterror}{\SI{0.091(3)}{\percent}} % Calculated as the error from the combination of the parity pulse on the network qubit followed by fluoresence detection
\newcommand{\bobmidcircuiterror}{\SI{0.122(2)}{\percent}} % Calculated as the error from the combination of the parity pulse on the network qubit followed by fluoresence detection

\newcommand{\aliceclockrotation}{\SI{0.016(1)}{\percent}} % 
\newcommand{\bobclockrotation}{\SI{0.015(1)}{\percent}} % 

\newcommand{\totalerror}{\SI{11.9(6)}{\percent}} % Calculated as the product of (1 - error), rather than sum since approximating as adding breaks down a bit

\newcommand{\quadrupolewavelength}{\SI{674}{\nano\meter}}
\newcommand{\ramanwavelength}{\SI{402}{\nano\meter}}
\newcommand{\hyperfinesplitting}{\ish\SI{3.2}{\giga\hertz}}
\newcommand{\bfield}{$\ish\SI{0.5}{\milli\tesla}$}
\newcommand{\clocksensitivity}{\SI{122}{\kilo\hertz\per\milli\tesla}}

\newcommand{\cyclelength}{\SI{1168}{\nano\second}} % 64 ns comes from handshake iirc
\newcommand{\tryduration}{\SI{200}{\micro\second}}
\newcommand{\triesperherald}{\num{7084}} % Includes protocol aborts
\newcommand{\successprobability}{\num{1.41e-4}} % 1 / \triesperherald{}
\newcommand{\betweeninteractions}{\SI{2.254}{\milli\second}} % Rounded from 2203.672us + 50us extra slack
\newcommand{\colddoppler}{\SI{1.254}{\milli\second}} % Rounded from 1203.672us + 50us extra slack
\newcommand{\eitduration}{\SI{1}{\milli\second}} % 
\newcommand{\entanglementrate}{\SI{9.7}{\per\second}} %
\newcommand{\entanglementtime}{\SI{103}{\milli\second}} %
\newcommand{\statetomographyshots}{\num{2e5}}
\newcommand{\interpulsedelay}{\SI{7.4}{\milli\second}} %
\newcommand{\midcircuitmeasurementduration}{\SI{500}{\micro\second}} %

\newcommand{\alicespamcirdown}{8.5(9)}
\newcommand{\alicespamcirup}{6.4(6)}

\newcommand{\alicespamauxdown}{4.7(7)}
\newcommand{\alicespamauxup}{3.6(4)}

\newcommand{\alicespamnetdown}{2.6(6)}
\newcommand{\alicespamnetup}{7.8(8)}

\newcommand{\bobspamcirdown}{6.0(8)}
\newcommand{\bobspamcirup}{7.5(6)}

\newcommand{\bobspamauxdown}{3.2(7)}
\newcommand{\bobspamauxup}{5.3(5)}

\newcommand{\bobspamnetdown}{6.5(8)}
\newcommand{\bobspamnetup}{4.5(6)}

\newcommand{\avgspam}{\num{5.0(2)e-3}}

\newcommand{\alicesrfluoerror}{\num{6.6(1)e-4}}
\newcommand{\bobsrfluoerror}{\num{5.51(2)e-4}}

\newcommand{\dougal}{DM}
\newcommand{\peter}{PD}
\newcommand{\davidN}{DPN}
\newcommand{\ellis}{EMA}
\newcommand{\ayush}{AA}
\renewcommand{\beth}{BCN}
\newcommand{\raghu}{RS}
\newcommand{\gabriel}{GA}
\newcommand{\davidL}{DML}

\begin{document}
\title{\Title{}}

\author{D.~Main}
\email{dougal.main@physics.ox.ac.uk}
\author{P.~Drmota}
\author{D.~P.~Nadlinger}
\author{E.~M.~Ainley}
\author{A.~Agrawal}
\author{B.~C.~Nichol}
\author{R.~Srinivas}
\author{G.~Araneda}
\author{D.~M.~Lucas}

\affiliation{Department of Physics, University of Oxford, Clarendon Laboratory, Parks Road, Oxford OX1 3PU, United Kingdom}

\begin{abstract}\noindent
\Ac{DQC} combines the computing power of multiple networked quantum processing modules, enabling the execution of large quantum circuits without compromising on performance and connectivity~\cite{grover_quantum_1997, cirac_distributed_1999}.
Photonic networks are well-suited as a versatile and reconfigurable interconnect layer for \ac{DQC}; remote entanglement shared between matter qubits across the network enables all-to-all logical connectivity via \ac{QGT}~\cite{jiang_distributed_2007, monroe_large-scale_2014}.
For a scalable \ac{DQC} architecture, the \ac{QGT} implementation must be deterministic and repeatable; until now, there has been no demonstration satisfying these requirements.
We experimentally demonstrate the distribution of quantum computations between two photonically interconnected trapped-ion modules.
The modules are separated by $\ish\SI{2}{\meter}$, and each contains dedicated \network{} and \circuit{} qubits.
By using heralded remote entanglement between the \network{} qubits, we deterministically teleport a \acl{CZ} gate between two \circuit{} qubits in separate modules, achieving $\SI{86}{\percent}$ fidelity.
We then execute Grover's search algorithm~\cite{grover_fast_1996} -- the first implementation of a distributed quantum algorithm comprising multiple non-local two-qubit gates -- and measure a $\SI{71}{\percent}$ success rate.
Furthermore, we implement distributed \iswap{} and \swap{} circuits, compiled with 2 and 3 instances of \ac{QGT}, respectively, demonstrating the ability to distribute arbitrary two-qubit operations~\cite{vidal_universal_2004}.
As photons can be interfaced with a variety of systems, this technique has applications extending beyond trapped-ion quantum computers, providing a viable pathway towards large-scale quantum computing for a range of physical platforms.

\acresetall
\end{abstract}

\maketitle
\noindent

\section{Introduction}
\begin{figure*}[t]
    \centering
    \includegraphics[width=178mm]{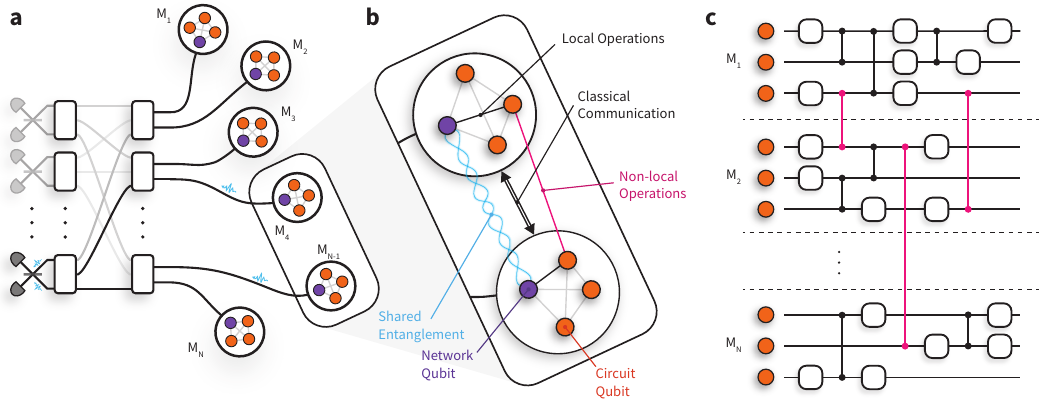}
    \caption{%
    \justifying
    \textbf{Distributed quantum computing architecture.}
    \textbf{a}, Schematic of a \acs*{DQC} architecture comprising photonically interconnected modules. Entanglement is heralded between \network{} qubits via the interference of photons on beamsplitters. A photonic switchboard provides a flexible and reconfigurable network topology.
    \textbf{b}, The modules consist of a small number of \network{} qubits (purple) and \circuit{} qubits (orange), which may directly interact via local operations. Quantum gate teleportation mediates non-local gate interactions (pink) between \circuit{} qubits in separate modules. These protocols require the resources of shared entanglement, local operations, and classical communication.
    \textbf{c}, A quantum circuit distributed across a network of small quantum processing modules that function together as a single, fully intra-connected quantum computer.
    }
    \label{fig:modulararchitecture}
\end{figure*}

The potential of quantum computing to revolutionise various fields ranging from cryptography to drug discovery is widely recognised~\cite{shor_polynomial-time_1997, cao_potential_2018}.
However, regardless of the physical platform used to realise the quantum computer, scaling up the number of qubits while maintaining precise control and inter-connectivity is a major technical challenge~\cite{bruzewicz_trapped-ion_2019, bravyi_future_2022, gill_quantum_2024}.
The \ac{DQC} architecture, depicted in \fig{fig:modulararchitecture}, addresses this challenge by enabling large quantum computations to be executed by a network of quantum processing modules~\cite{grover_quantum_1997, cirac_distributed_1999}.
The modules each host a relatively small number of qubits and are interconnected via both classical and quantum information channels.
By preserving the reduced complexity of the individual modules and transforming the scaling challenge into the task of building more modules and establishing an interface between them, the \ac{DQC} architecture provides a scalable approach to fault-tolerant quantum computing~\cite{jiang_distributed_2007, monroe_large-scale_2014}.

The interface between modules could be realised by directly transferring quantum information between modules.
However, losses in the interconnecting quantum channels would lead to the unrecoverable loss of quantum information.
Quantum teleportation offers a loss-less alternative interface, using only bipartite entanglement (e.g.\ Bell states) shared between modules, together with \ac{LOCC} to effectively replace the direct transfer of quantum information across quantum channels~\cite{bennett_teleporting_1993, gottesman_demonstrating_1999}.
\Ac{QGT} efficiently implements non-local entangling gates between qubits in separate modules, consuming only one Bell pair and the exchange of two classical bits~\cite{eisert_optimal_2000, collins_nonlocal_2001}, as depicted in \fig{fig:modulararchitecture}(b).
Given arbitrary single- and two-qubit operations within each node, \ac{QGT} completes a universal gate set for the distributed quantum computer~\cite{gottesman_demonstrating_1999}.
The primary advantage of teleportation-based schemes over direct transfer is the exclusive use of the quantum channel for generating identical Bell states; channel losses can be overcome by repetition without losing quantum information, and the distance between modules can be increased by inserting quantum repeaters~\cite{briegel_quantum_1998}.
Additionally, channel noise may be suppressed using entanglement purification~\cite{dur_entanglement_2003}.
Since teleportation protocols are executed strictly after entanglement has been established, they enable continuous deterministic operation even if the entanglement is generated non-deterministically.
This deterministic nature is crucial for the scalability, eliminating the need for post-selection of singular successful outcomes out of an exponentially large set of undesired outcomes.
Teleportation protocols are agnostic to the physical implementation of the quantum channels, making them a versatile tool for \ac{DQC} across different platforms.
In the trapped-ion \ac{QCCD} architecture, qubits can be dynamically transported between modules within a single chip~\cite{kielpinski_architecture_2002} -- or even across chips~\cite{akhtar_high-fidelity_2023} -- and thus be used to mediate entangling gates between different trap zones~\cite{wan_quantum_2019, pino_demonstration_2021}.
Photons, however, make natural carriers of quantum information since they can travel long distances without significant degradation of their quantum state.
Photonic interconnects enable all-to-all connectivity between qubits distributed across the network whose topology can be dynamically reconfigured without the need to open up complex vacuum and/or cryogenic systems.
Moreover, optical components are widely available and can be operated under ambient conditions.
These properties make photonic interconnects particularly appealing for networking quantum computing modules, as shown in \fig{fig:modulararchitecture}(a).
As depicted in \fig{fig:modulararchitecture}(b), we consider modules containing ``\network{}" and ``\circuit{}" qubits with full interconnectivity via local quantum operations.
Remote entanglement of \network{} qubits in separate modules is generated by the interference of photons, where reconfigurability and flexibility could be provided via a photonic switchboard.
This entanglement can then be used to mediate entangling gates between the \circuit{} qubits in different modules via \ac{QGT}, enabling the network to function as a single fully connected quantum processor, as shown in \fig{fig:modulararchitecture}(c).
Heralded entanglement between spatially separated qubits has been achieved experimentally in a variety of platforms including diamond colour centres~\cite{humphreys_deterministic_2018, knaut_entanglement_2024}, superconducting qubits~\cite{storz_loophole-free_2023}, neutral atoms~\cite{ritter_elementary_2012, van_leent_entangling_2022}, and trapped-ions~\cite{moehring_entanglement_2007, stephenson_high-rate_2020, saha_high-fidelity_2024}.
\Ac{QGT} has been implemented probabilistically in purely photonic systems, requiring passive optical elements and post-selection to perform the conditional rotations that complete the gate teleportation~\cite{huang_experimental_2004, gao_teleportation-based_2010}.
\authorcite{chou_deterministic_2018} demonstrated deterministic teleportation of a controlled-NOT gate between two qubits encoded in the modes of two superconducting cavities on the same device, separated by $\ish\SI{2}{\centi\meter}$, while a third cavity enabled the deterministic generation of entanglement between two transmon \network{} qubits.
Recently, there have been demonstrations of \ac{QGT} between superconducting qubits within a single device, demonstrating the viability of \ac{QGT} to overcome nearest neighbour constraints in this architecture~\cite{baumer_efficient_2023, hashim_efficient_2024}.
In the trapped-ion \ac{QCCD} architecture, \authorcite{wan_quantum_2019} demonstrated \ac{QGT} in which the entanglement was deterministically generated between two ``\network{}" qubits via local operations before being transported $\ish\SI{840}{\micro\meter}$ to two separate locations within the same trap.
\authorcite{daiss_quantum-logic_2021} demonstrated a heralded non-local entangling gate across a photonic quantum network using a photon to directly transfer quantum information between modules.
However, photon loss necessarily destroys the states of the \circuit{} qubits, rendering this scheme non-deterministic.
Until now, there has been (i) no demonstration of deterministic \ac{QGT} across a quantum network,
and (ii) no demonstration of distributed circuits comprising multiple non-local entangling gates.
In photonic platforms, this has been prevented by the inability to store the photons between interactions~\cite{huang_experimental_2004, gao_teleportation-based_2010}, while in the \ac{QCCD} demonstration, this was limited by the decoherence of the \circuit{} qubits during the generation of entanglement~\cite{wan_quantum_2019}.

\begin{figure*}[t]
    \centering
    \includegraphics[width=178mm]{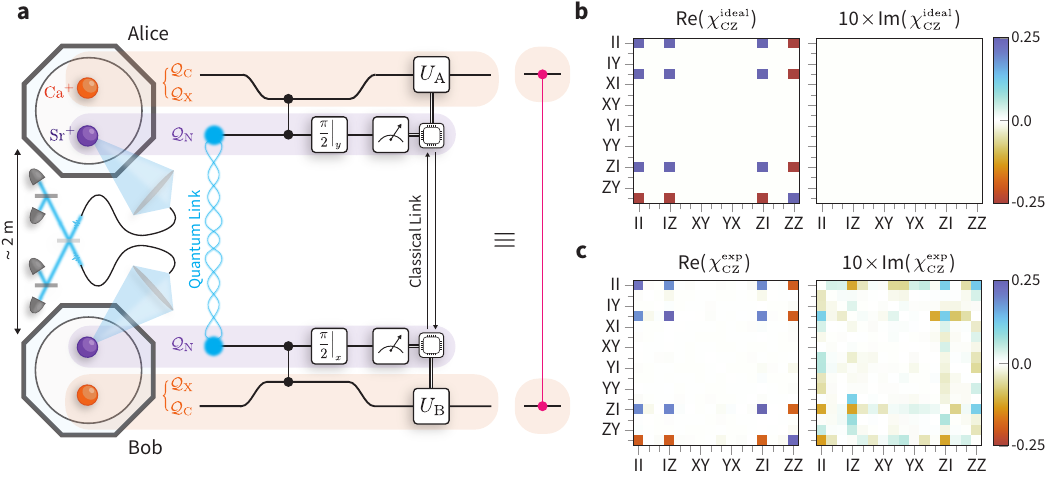}
    \caption{%
    \justifying
    \textbf{Teleportation of a \ac*{CZ} gate between two trapped-ion modules.}
    \textbf{a}, The two modules, Alice and Bob, each hold a \Sr{88} ion (purple) and a \Ca{43} ion (orange). \Sr{} provides a \network{} qubit, $\Qnet$, while \Ca{} provides both a long-lived \circuit{} qubit, $\Qcir$, and an auxiliary qubit, $\Qaux$. Prior to the protocol, the \circuit{} qubits are in some arbitrary state. The protocol begins by generating entanglement between the \network{} qubits via a photonic link. Upon heralding entanglement, each module applies a local \acs*{CZ} gate between the \network{} and \circuit{} qubits, using the \auxiliary{} qubit temporarily to mediate the gate mechanism. The outcomes of mid-circuit parity measurements of the \network{} qubits are exchanged in real-time via a classical (TTL) link connecting the control systems of the two modules. This information is used to condition local feed-forward operations, $U_\mathrm{A}$ and $U_\mathrm{B}$, on the \circuit{} qubits -- completing the teleportation of the \acs*{CZ} gate. 
    \textbf{b}, Process matrix for an ideal \acs*{CZ} gate. 
    \textbf{c}, Measured process matrix, reconstructed via \acl*{QPT}, yielding an average gate fidelity of \czfidelity{}, compared to an ideal \acs*{CZ} gate.
    }
    \label{fig:czteleportation}
\end{figure*}

In this work, we present the first demonstration of \ac{DQC} across a network of two trapped-ion modules, each containing a \network{} qubit and a \circuit{} qubit, and separated by a macroscopic distance ($\ish\SI{2}{\meter}$).
We mediate deterministic two-qubit \ac{CZ} interactions between the \circuit{} qubits via \ac{QGT}, utilizing entanglement previously established across the network between the two \network{} qubits.
Leveraging the robust storage of quantum information in the \circuit{} qubits while generating subsequent rounds of entanglement between \network{} qubits~\cite{drmota_robust_2023}, we execute distributed quantum circuits comprising multiple non-local two-qubit gates.
We demonstrate the distributed \iswap{} and \swap{} gates, which consist of 2 and 3 instances of \ac{QGT}, respectively.
The actions of all teleported gates are characterized using \ac{QPT}.
Finally, we implement Grover's algorithm on our distributed quantum computer.

\section{Teleportation of a Controlled-Z Gate}
Our apparatus, depicted in \fig{fig:czteleportation}(a), consists of two trapped-ion modules, Alice and Bob, each co-trapping one \Sr{88} ion and one \Ca{43} ion~\methods{methods:apparatus}.
The \Ca{} ion provides a magnetic field-insensitive ``\circuit{}" qubit, $\Qcir:=\{\ket{\down{\cir{}}}\equiv\ket{F=4, m_F=0}, \ket{\up{\cir{}}}\equiv\ket{F=3, m_F=0}\}$, in the ground hyperfine manifold which has been used to demonstrate state-of-the-art quantum logic~\cite{harty_high-fidelity_2016, ballance_high-fidelity_2016}.
The \Sr{} ion, on the other hand, provides an efficient interface to the optical quantum network~\cite{stephenson_high-rate_2020}.
We define the \network{} qubit in \Sr{} by $\Qnet=\{\ket{\down{\net{}}}\equiv\ket{\S{1/2}, m_J=-\frac{1}{2}}, \ket{\up{\net{}}}\equiv\ket{\D{5/2}, m_J=-\frac{3}{2}}\}$.
To implement local entangling operations between these two species, we employ the light shift gate mechanism~\cite{hughes_benchmarking_2020} between $\Qnet$ and an auxiliary qubit in the ground hyperfine manifold of \Ca{}, $\Qaux{} := \{\ket{\down{\aux{}}}\equiv\ket{F=4, m_F=+4}, \ket{\up{\aux{}}}\equiv\ket{F=3, m_F=+3}\}$, which, unlike the $\Qcir$ qubit, experiences the necessary light shifts~\methods{methods:localgates}.
At the points at which we want to perform the local entangling gate, we transfer the quantum information stored in $\Qcir{}$ temporarily to $\Qaux{}$ to perform the gate operations~\methods{methods:hyperfinetransfer}.
The \ac{QGT} protocol used here to mediate \ac{CZ} gates between the \circuit{} qubits in separate modules is depicted in \fig{fig:czteleportation}.
We allow the \circuit{} qubits to start in an arbitrary state $\ket{\psi^{\mathrm{AB}}_\mathrm{in}}\in\Qcir^{\otimes2}$, which could be part of a larger, long-running computation.
We begin the \ac{QGT} protocol by generating the remotely entangled Bell state,
\[
\ket{\Psi^{+}}=\frac{\ket{\up{}\down{}}+\ket{\down{}\up{}}}{\sqrt{2}}\in\Qnet^{\otimes2},
\]
between the \network{} qubits~\cite{stephenson_high-rate_2020}, with a fidelity of \rawfidelity{}~\methods{methods:remoteentanglement}.
This is done via a try-until-success process, where a herald indicates a success.
The \circuit{} qubits provide a robust quantum memory~\cite{drmota_robust_2023}, enabling storage of the encoded quantum information until the remote entanglement is successfully heralded.
At this stage, we map the state stored in the \circuit{} qubits ($\Qcir$) to the auxiliary qubits ($\Qaux$) in preparation for the local entangling operations~\methods{methods:hyperfinetransfer}.
In each module, we perform local \ac{CZ} gates between the \network{} and \auxiliary{} qubits~\methods{methods:localgates}, before transferring the \auxiliary{} qubit back to the \circuit{} qubit.
We then perform mid-circuit measurements of the \network{} qubits in the $X$ and $Y$ bases in Alice and Bob, respectively.
The modules exchange the measurement outcomes in real-time -- using a classical (TTL) link between their control systems -- and perform single-qubit feed-forward operations conditioned on the exchanged bits to complete the gate teleportation protocol~\methods{methods:conditionaloperations}. This implements the non-local gate $\ket{\psi^{\mathrm{AB}}_\mathrm{in}}\rightarrow U_{\mathrm{CZ}}^{\mathrm{AB}}\ket{\psi^{\mathrm{AB}}_\mathrm{in}}$.
We characterise the \ac{QGT} protocol using \ac{QPT} to reconstruct the process matrix, $\chi^{\mathrm{exp}}_\mathrm{CZ}$, providing a complete description of the action of the teleported \ac{CZ} gate on the two \circuit{} qubits.
Compared to the ideal \ac{CZ} process, shown in \fig{fig:czteleportation}(b), the reconstructed process matrix for the teleported gate, shown in \fig{fig:czteleportation}(c), has an average gate fidelity of \czfidelity{}.
The \ac{QGT} protocol is completely self-contained -- the input states of the \circuit{} qubits are set prior to the execution of the non-local gate -- and output states are available for further computation.
With single-qubit rotations of the \circuit{} qubits, this teleported \ac{CZ} gate is a key element of a gate set for \ac{DQC}, enabling the modules to act as a single, fully-connected universal quantum processor.

\section{Distributed Quantum Computing}
\begin{figure*}[t]
    \centering
    \includegraphics[width=178mm]{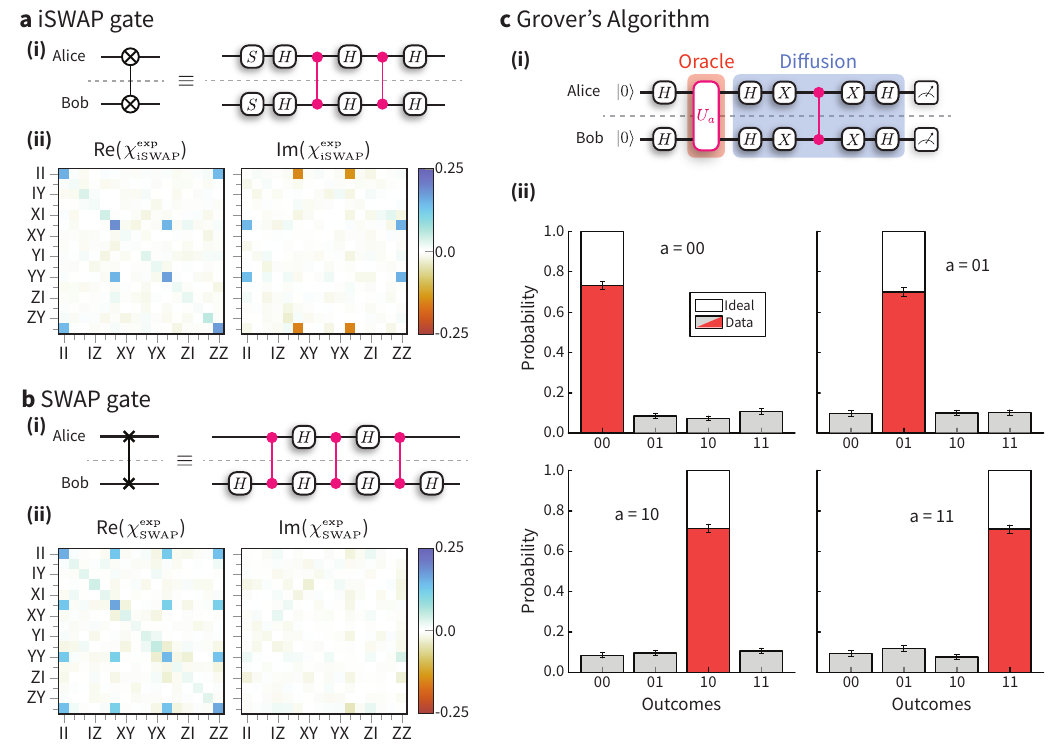}
    \caption{%
    \justifying
    \textbf{Distributed quantum computing results.}
    \Ac*{CZ} decompositions of the distributed \textbf{a}(i) \iswap{} and \textbf{b}(i) \swap{} circuits, comprising two and three instances of \ac*{QGT}, respectively. The reconstructed process matrices for the \textbf{a}(ii) \iswap{} and \textbf{b}(ii) \swap{} gates indicate an average gate fidelity of \iswapfidelity{} and \swapfidelity{}, respectively.
    \textbf{c}, Grover's algorithm: (i) circuit comprising two instances of \ac*{QGT}: the first implements the Grover oracle call which marks a particular state, $a$, while the second implements the diffusion circuit, (ii) measurement outcomes from 500 repetitions of Grover's algorithm per marked state; the average success probability is \groveraverage{}.}
    \label{fig:distributedcircuits}
\end{figure*}

In general, any arbitrary two-qubit unitary operation can be decomposed into at most three \ac{CZ} gates~\cite{vidal_universal_2004}.
We demonstrate our ability to perform sequential rounds of \ac{QGT} by executing the \ac{CZ} decompositions of the \iswap{} and \swap{} gates, shown in Figs. \ref{fig:distributedcircuits}(a)(i) and \ref{fig:distributedcircuits}(b)(i), comprising two and three instances of \ac{QGT}, respectively.
As with the teleported \ac{CZ}, we characterise these circuits via \ac{QPT}, see Figs. \ref{fig:distributedcircuits}(a)(ii) and \ref{fig:distributedcircuits}(b)(ii).
From the reconstructed process matrices, we measure average gate fidelities of \iswapfidelity{} and \swapfidelity{} for the \iswap{} and \swap{} gates, respectively.
By constructing circuits with multiple instances of \ac{QGT} -- enabled by our ability to perform \ac{QGT} deterministically and on-demand -- we demonstrate the ability to perform universal \ac{DQC}.

Finally, we implement Grover's algorithm~\cite{grover_fast_1996} on our distributed quantum processor. 
This algorithm considers searching through a set of unsorted items, $x\in L$, to find a particular item, $a\in L$. 
The search problem is represented by the function
\[
f_a(x)=\begin{cases}
    1 & \text{if } x = a,\\
    0 & \text{otherwise.}
\end{cases}
\]
In the two-qubit case, there are four items to search through.
Classically, the item $a$ could be identified with, on average, two queries of the function $f_a(x)$.
Using the quantum circuit shown in \fig{fig:distributedcircuits}(c)(i), the same task can be accomplished with only one query.
After preparing a superposition of all possible inputs with parallel Hadamard gates, an instance of \ac{QGT} implements the \textit{oracle}, which performs the mapping $U_a: \ket{x}\rightarrow(-1)^{f_a(x)}\ket{x}$, marking the state $\ket{a}$.
A second instance of \ac{QGT} implements the Grover \textit{diffusion} circuit, which decodes the quantum information provided by the oracle into an observable.
In the two-qubit case considered here, the application of the Grover diffusion circuit should leave the register in the state $\ket{a}$, which is the solution to the function $f_a$, and thus a measurement of the register yields the solution to the search problem with unit probability.
In the case of $N$ items, to approach unit probability of obtaining the solution, one would require $\ish\mathcal{O}(\sqrt{N})$ iterations of the oracle-diffusion circuit, compared with $\ish\mathcal{O}(N/2)$ for a classical search.
The results of Grover's algorithm -- executed on our distributed quantum processor -- are shown in \fig{fig:distributedcircuits}(c)(ii).
For the marked states $a\in\{00, 01, 10, 11\}$, we obtain the correct result with an average success rate of \groveraverage{}.
To our knowledge, this represents the first deterministic execution of any algorithm on a distributed quantum computer.

\section{Discussion}
The performance of our distributed quantum circuits is consistent with the errors from the teleported \ac{CZ} gates.
We summarise the leading error sources affecting our teleported \ac{CZ} gate in \tab{tab:errorbudget}.
The measured fidelity of our gate is slightly lower than that predicted by the error budget, which we attribute to drifts in the calibration of various components over the duration of the data acquisition.
The majority of identified error sources occur during local operations in each module. 
It is worth noting that our local errors do not represent the state-of-the-art of trapped-ion processors; however, local operations exceeding the fidelity threshold for fault-tolerant quantum computing have been demonstrated in this platform~\cite{ballance_high-fidelity_2016, hughes_benchmarking_2020, srinivas_high-fidelity_2021, moses_race-track_2023, weber_robust_2024}.
Relevant to our implementation, \authorcite{hughes_benchmarking_2020} demonstrated mixed-species two-qubit gates between \Sr{88} and \Ca{43} ions with a gate error of $\SI{0.2(1)}{\percent}$.
We therefore conclude that the technical limitations in our implementation can be overcome.
The other significant source of error is the remote entanglement of the \network{} qubits across the photonic quantum network; we observe a fidelity of the remotely entangled \network{} qubits to the desired $\ket{\Psi^+}$ state of \rawfidelity{}.
Unlike the local operations, the performance of our remote entanglement is at the state-of-the-art.
To improve this, and hence enable the teleportation of high-fidelity entangling gates between modules, entanglement purification could be used to distribute high-fidelity entangled states from a number of lower-fidelity entangled states~\cite{dur_entanglement_2003, nigmatullin_minimally_2016}.

\begin{table}[ht]
    \centering
    \begin{tblr}{
        colspec = {ccc},
        cell{1}{2} = {c=2}{c}, % multicolumn (column number=2, center alignment)
        cell{3}{2} = {c=2}{c},
        cell{9}{2} = {c=2}{c},
    }
    \hline[2pt]
    Source &   Error &       \\
    \hline[1pt]
    &    Alice &  Bob \\
    \cline{2-5}
    Raw entanglement  & \rawinfidelity &  \\
    Mixed-species gate  &  \alicewzz & \bobwzz  \\
    $\Q{C}$ decoherence & \alicememoryerrors &  \bobmemoryerrors \\
    $\Qaux \leftrightarrow \Qcir$ transfer  &   \alicetransfer &  \bobtransfer  \\
    Mid-circuit measurement &   \alicemidcircuiterror &  \bobmidcircuiterror  \\
    $\Qcir{}$ rotations &   \aliceclockrotation &  \bobclockrotation  \\
    \cline{2-5}
    Predicted total error & \totalerror &\\
    \hline[2pt]
\end{tblr}

    \caption{%
    \justifying
    \textbf{Error budget for \ac{CZ} gate teleportation.} 
    The characterisation of each error contribution is discussed in~\methods{methods:apparatus}.
    }
    \label{tab:errorbudget}
\end{table}

Our implementation features a single \circuit{} qubit in each module; however processors with larger numbers of qubits have been realised.
With only 3 \circuit{} qubits (and one \network{} qubit) per module, the purification of arbitrary quantum channels would be possible~\cite{nigmatullin_minimally_2016}.
The capabilities of the individual modules may be extended even further by deploying the \ac{QCCD} architecture.
With recent demonstrations in both academic research~\cite{hilder_fault-tolerant_2022} and industry~\cite{pino_demonstration_2021} highlighting the power of this approach, embedding these systems in a quantum network would combine their power with the reconfigurability and flexibility of the \ac{DQC} architecture.
Conversely, computational bottlenecks associated with ion transport overheads observed in the \ac{QCCD} architecture~\cite{pino_demonstration_2021} could be mitigated using photonic interconnects integrated into a single device~\cite{knollmann_integrated_2024}.
While the results presented here were achieved using trapped-ion quantum processing modules, photons may be interfaced with a variety of systems.
The connectivity and reconfigurability enabled by photonic networks provides a scalable approach for other quantum computing platforms such as diamond colour centres and neutral atoms.
Additionally, modules of different platforms could be connected via wavelength conversion, enabling a hybrid \ac{DQC} platform.
Furthermore, teleportation protocols are not limited to qubits; they can be extended to higher-dimensional quantum computing paradigms, such as qudits~\cite{luo_quantum_2019} and \ac{CVQC}~\cite{lloyd_quantum_1999, walshe_continuous-variable_2020}, allowing these platforms to benefit from the \ac{DQC} architecture.
Quantum repeater technology~\cite{briegel_quantum_1998} would enable large physical separation between the quantum processing modules, thereby paving the way for the development of a \qinternet{}~\cite{wehner_quantum_2018}.
The scope of these networks extends beyond quantum computing technologies; the ability to control distributed quantum systems, as enabled by this architecture, to engineer complex quantum resources has applications in multi-partite secrete sharing~\cite{hillery_quantum_1990}, metrology~\cite{komar_quantum_2014}, and probing fundamental physics~\cite{greenberger_bells_1990}.
\section{Methods}
\subsection{Dual-species ion-trap modules}\label{methods:apparatus}
Our apparatus comprises two trapped-ion processing modules, Alice and Bob.
Each module, depicted in \extfig{extfig:apparatus}, consists of an ultra-high vacuum chamber containing a room-temperature, micro-fabricated surface Paul trap; the trap used in Alice (Bob) is a \hoa{}~\cite{maunz_high_2016} (\phoenix{}~\cite{revelle_phoenix_2020}) trap, fabricated by Sandia National Laboratories.
In each module, we co-trap \Sr{88} and \Ca{43} ions.
Each species of ion is addressed via a set of lasers used for cooling, state-preparation, and readout.
A high-numerical aperture (0.6 NA) lens enables single-photon collection from the \Sr{} ions.
A \bfield{} magnetic field is applied parallel to the surface of the trap to provide a quantisation axis.
As outlined in the main text, the \Sr{} ion provides an optical \network{} qubit, $\Qnet{}$, which is manipulated directly using a \quadrupolewavelength{} laser.
The ground hyperfine manifold of the \Ca{} ion provides a \circuit{} qubit, $\Qcir{}$.
At \bfield{}, the sensitivity of the $\Qcir{}$ qubit transition frequency to magnetic field fluctuations is \clocksensitivity{}, i.e. $\ish 2$ orders of magnitude lower than that of the $\Qnet$ qubit -- making it an excellent memory for quantum information~\cite{drmota_robust_2023}.
Additionally, we define an auxiliary qubit, $\Qaux{}$, in the ground hyperfine manifold of \Ca{} for implementing local entangling operations, state preparation, and readout.
The measured \ac{SPAM} errors for each qubit are presented in \extfig{exttab:spam}. 
The spectral isolation between the two species allows us to address one species without causing decoherence of the quantum information encoded in the other species.
We make use of this for sympathetic cooling, mid-circuit measurement, and interfacing with the quantum network during circuits.

\subsection{Quantum process tomography}\label{methods:qpt}
The action of a quantum process acting on a system of $N$ qubits may be represented by the process matrix $\chi_{\alpha\beta}$ such that,
\begin{equation*}
    \mathcal{E}\left(\rho\right)=\sum_{\alpha,\beta=0}^{D-1}\chi_{\alpha\beta} P_\alpha \rho P_\beta^\dagger,
\end{equation*}
where $P_\alpha \in\mathcal{P}^{\otimes N}$ are the set of single-qubit Pauli operators $\mathcal{P}=\left\{\mathbb{I}, \sigma_x, \sigma_y, \sigma_z\right\}$, and $D=\mathrm{dim}\left(\mathcal{P}^{\otimes N}\right)=4^N$.
\Ac{QPT} enables us to reconstruct the matrix $\chi_{\alpha\beta}$, thereby attaining a complete characterisation of the process.
\Ac{QPT} is performed by preparing the qubits in the states $\rho_i=\ket{\psi_i}\bra{\psi_i}$, where $\ket{\psi_i}$ are chosen from a tomographically complete set
\begin{equation}\label{methodeq:tomo-states}
    \ket{\psi_i}\in\left\{\ket{\down{}}, ~\ket{\up{}},~\frac{\ket{\down{}}+\ket{\up{}}}{\sqrt{2}}, ~\frac{\ket{\down{}}+\mathrm{i}\ket{\up{}}}{\sqrt{2}}\right\},
\end{equation}
performing the process, $\mathcal{E}$, followed by measuring the output state $\mathcal{E}[\rho_i]$ in a basis chosen from a tomographically complete set.
Using diluted maximum-likelihood estimation~\cite{rehacek_diluted_2007}, the outcomes of the measurements can be used to reconstruct the $\chi$-matrix representing the process.
In practice, the input states are created by rotating $\ket{0}$ to $\ket{\psi_i} = R_i \ket{0}$ with
\begin{align}\label{eq:tomographic rotations}
    R_i \in \left\{
        \mathbb{I},
        \sigma_x,
        \tfrac{1}{\sqrt{2}}\left(\mathbb{I} - \mathrm{i} \sigma_y\right),
        \tfrac{1}{\sqrt{2}}\left(\mathbb{I} + \mathrm{i} \sigma_x \right)
    \right\} .
\end{align}
Likewise, the tomographic measurements are performed by rotating the output state $\mathcal{E}[\rho_i]$ by $R_j$ [Eq.~\eqref{eq:tomographic rotations}], and subsequently measuring it in the $\sigma_z$ basis.
Ideally, this sequence implements the projectors $P_{0,j} = \ket{\psi_j}\bra{\psi_j}$ and $P_{1,j} = \ket{\psi_{\perp, j}}\bra{\psi_{\perp, j}}$, where $\bra{\psi_{\perp, j}}\ket{\psi_j} = 0$.
However, \ac{SPAM} errors would manifest as errors in the reconstructed process; we therefore model these errors by replacing the $\sigma_z$ measurement with \acp{POVM},
\begin{align*}
    M_0&=\left(1-\epsilon_{\ket{0}}\right)\ket{0}\bra{0}+\epsilon_{\ket{1}}\ket{1}\bra{1}\\
    M_1&=\left(1-\epsilon_{\ket{1}}\right)\ket{1}\bra{1}+\epsilon_{\ket{0}}\ket{0}\bra{0},
\end{align*}
where $\epsilon_{\ket{0}}$ ($\epsilon_{\ket{1}}$) is the \ac{SPAM} error associated with the $\ket{0}$ ($\ket{1}$) qubit state.
The values used for these operators are given in \extfig{exttab:spam}.
Resampling of the measurement outcomes is used to generate new data sets, which are analysed in the same way as the original data set, and are used to determine the sensitivity of the analysis to the statistical fluctuations in the input data.
Error bars on the fidelities of reconstructed processes are quoted as the standard deviation of the fidelities of the resampled data sets.

\subsection{Remote entanglement generation}\label{methods:remoteentanglement}
The heralded generation of remote entanglement between \network{} qubits in separate modules, outlined in \authorcite{stephenson_high-rate_2020}, is central to our \ac{QGT} protocol.
Photons entangled with the \Sr{} ions are collected from each module using high-numerical aperture lenses and single-mode optical fibres bring the photons to a central Bell state analyser, where a measurement of the photons projects the ions into a maximally entangled state.
This forms the photonic quantum channel interconnecting the two modules.
Following \authorcite{stephenson_high-rate_2020}, we use a \SI{674}{\nano\meter} $\pi$-pulse to map the remote entanglement from the ground-state Zeeman qubit to an optical qubit, which we refer to as the \network{} qubit, to minimise the number of quadrupole pulses in subsequent operations.
Successful generation of entanglement is heralded by particular detector click patterns and, after subsequent local rotations, indicates the creation of the maximally entangled $\Psi^+$ Bell-state,
\[
\ket{\Psi^{+}}=\frac{\ket{\up{}\down{}}+\ket{\down{}\up{}}}{\sqrt{2}}\in\Qnet^{\otimes2}.
\]
This process is executed while simultaneously storing quantum information in the \circuit{} qubits which, as demonstrated by \authorcite{drmota_robust_2023}, are robust to this network activity.
Each entanglement generation attempt takes \cyclelength{} and it takes \triesperherald{} attempts to successfully herald entanglement on average, corresponding to a success probability of \successprobability{}.
To mitigate heating of the ion-crystal, we interleave \tryduration{} of entanglement generation attempts with \betweeninteractions{} of sympathetic re-cooling of the \Sr{}-\Ca{} crystal using the \Sr{} ion.
The sympathetic recooling comprises \colddoppler{} of Doppler cooling, followed by \eitduration{} of \ac{EIT} cooling.
Overall, this results in an average entanglement generation rate of \entanglementrate{} (equivalently, it takes on average \entanglementtime{} to generate entanglement between \network{} qubits), although this rate could be increased by optimising the interleaved cooling sequence.
This rate is lower than the \SI{182}{\per\second} rate previously reported in our apparatus~\cite{stephenson_high-rate_2020} due to the additional cooling.
We characterise the remote entanglement using quantum state tomography; by performing tomographic measurements on \statetomographyshots{} copies of the remotely entangled state, we reconstruct the density matrix of the \network{} qubits, $\rho^{AB}_N$, shown in \extfig{extfig:entanglementgeneration}(d).
In order to isolate the fidelity of the \enquote{Quantum Link} in Fig.~\ref{fig:czteleportation}, we account for the imperfect tomographic measurements in the reconstruction of the density matrix.
The fidelity of the reconstructed state to the desired $\Psi^+$ Bell-state, given by $\bra{\Psi^+}\rho^{AB}_N\ket{\Psi^+}$, is \rawfidelity{}.

\subsection{Circuit qubit memory during entanglement generation}\label{methods:memoryerrors}
Since each instance of \ac{QGT} requires the generation of entanglement between \network{} qubits, it is necessary to ensure that the \circuit{} qubits preserve their encoded quantum information during this process.
Due to their low sensitivity to magnetic field fluctuations, the \circuit{} qubits have exhibited $\ish\SI{100}{\milli\second}$ coherence times, and in previous work we demonstrated these qubits to be robust to network activity~\cite{drmota_robust_2023}.
We further suppress dephasing through dynamical decoupling.
Typically, dynamical decoupling is implemented over a fixed period of time; however the success of the entanglement generation process is non-deterministic and would therefore leave the dynamical decoupling sequence incomplete.
One solution would be to complete the dynamical decoupling pulse sequence once the entanglement has been generated, however it is desirable to minimise the time between heralding the entanglement generation and performing the \ac{QGT} protocol, in order to prevent dephasing of the \network{} qubits.
Instead, we make use of the fact that the action of a dynamical decoupling pulse on one of the \circuit{} qubits can be propagated through the teleported \ac{CZ} gate as
\begin{equation}\label{methodeq:ddcorrection}
    (X\otimes I)U_{\mathrm{CZ}} = U_{\mathrm{CZ}}(X\otimes Z).
\end{equation}
We therefore perform the dynamical decoupling pulses on the \circuit{} qubits until we obtain a herald of remote entanglement, at which point we immediately perform the \ac{QGT} sequence -- implementing a \ac{CZ} gate on the state of the \circuit{} qubits at the point of interruption.
Once this gate is completed, we perform the remaining dynamical decoupling pulses (without any inter-pulse delay), and use \eqref{methodeq:ddcorrection} to apply the appropriate $Z$ rotations required to correct for the propagation through the \ac{CZ} gate.
With this method, we suppress the dephasing errors in the \circuit{} qubits during entanglement generation, while minimising the time between successfully heralding the entanglement and consuming it for \ac{QGT}.
We deploy Knill dynamical decoupling~\cite{wang_single_2021, souza_robust_2011} with a \interpulsedelay{} inter-pulse delay (corresponding to a pulse every three rounds of interleaved entanglement attempts and re-cooling).
We use \ac{QPT} to reconstruct the process of storing the quantum information while generating entanglement; ideally, this process would not alter the quantum information stored in the \circuit{} qubit.
\Ac{QPT} is implemented by choosing input states for the \circuit{} qubits from the tomographically complete set given in \eqref{methodeq:tomo-states}, generating remote entanglement between the \network{} qubits while dynamically decoupling the \circuit{} qubits, then upon successful herald, completing the dynamical decoupling sequence and performing tomographic measurements of the \circuit{} qubits.
The reconstructed process matrices for each module corresponding to the action of storing quantum information during entanglement generation are shown in \extfig{extfig:entanglementgeneration}(c).
We observe fidelities to the ideal operation of \alicememoryfidelity{} and \bobmemoryfidelity{} for Alice and Bob, respectively.

\subsection{Local mixed-species entangling gates}\label{methods:localgates}
The ability to perform logical entangling gates between ions of different species allows us to separate the roles of \network{} and \circuit{} ions.
We implement mixed-species entangling gates following the approach taken by \authorcite{hughes_benchmarking_2020}, in which geometric phase gates are deterministically executed using a single pair of \SI{402}{\nano\meter} Raman beams, as depicted in \extfig{extfig:mixedspeciesgates}.
Here, we apply the gate mechanism directly to the \network{} qubit in \Sr{} -- rather than the Zeeman ground state qubit, as done in \authorcite{hughes_benchmarking_2020} and \authorcite{drmota_robust_2023} -- at the cost of a slightly reduced gate efficiency that is compensated for by the use of higher laser powers.
This enables us to perform mixed-species \ac{CZ} gates between the \network{} and \auxiliary{} qubits.
We characterise our mixed-species entangling gates using \ac{QPT} in each module, reconstructing the process matrices $\chi_{\text{CZ}}$ representing the action of the local \ac{CZ} gate acting between the \network{} and \auxiliary{} qubits.
The reconstructed process matrices for each module are shown in \extfig{extfig:mixedspeciesgates}(d).
Compared to the ideal \ac{CZ} gate, we observe average gate fidelities of \alicewzzfidelity{} and \bobwzzfidelity{} for Alice and Bob, respectively.

\subsection{Hyperfine qubit transfer}\label{methods:hyperfinetransfer}
Since the circuit qubit does not participate in the mixed-species gate, the gate interaction is performed on the network and auxiliary qubits.
Consequently, we require the ability to map coherently between the circuit and auxiliary qubit before and after the local operations.
As shown in \extfig{extfig:hyperfinetransfer}, this mapping is performed using a pair of \ramanwavelength{} Raman beams detuned by \hyperfinesplitting{}, to coherently drive the transitions within the ground hyperfine manifold of \Ca{}.
The transfer of the \circuit{} qubit to the \auxiliary{} qubit begins with the mapping of the state $\ket{\down{\cir{}}}$ to the state $\ket{\down{\aux{}}}$.
However, due to the near degeneracy of the transition $\mathcal{T}_0: \ket{\down{\cir{}}}\leftrightarrow\ket{F=3, M_F=+1}$ and the transition $\mathcal{T}_1:\ket{\up{\cir{}}}\leftrightarrow\ket{F=4, M_F=+1}$ (see \extfig{extfig:hyperfinetransfer}), separated by only $\ish\SI{15}{\kilo\hertz}$, it is not possible to map the $\ket{\down{\cir{}}}$ state out of the \circuit{} qubit without off-resonantly driving population out of the $\ket{\up{\cir{}}}$ state.
We suppress this off-resonant excitation using a composite pulse sequence, shown in \extfig{extfig:hyperfinetransfer}(b)(i), comprising three pulses resonant with the $\mathcal{T}_0$ transition, with pulse durations equal to the $2\pi$-time of the $\mathcal{T}_1$ transition, and phases optimised to minimise the off-resonant excitation.
This pulse sequence allows us to simultaneously perform a $\pi$-pulse on the $\mathcal{T}_0$ transition and the identity on the off-resonantly-driven $\mathcal{T}_1$ transition.
Raman $\pi$-pulses are then used to complete the mapping to the $\ket{\down{\aux{}}}$ state.
Another sequence of Raman $\pi$-pulses coherently maps $\ket{\up{\cir{}}}\rightarrow\ket{\up{\aux{}}}$, thereby completing the transfer of the \circuit{} qubit to the \auxiliary{} qubit, $\Qcir{}\rightarrow\Qaux{}$.
To implement the mapping $\Qaux{}\rightarrow\Qcir{}$, the same pulse sequence is applied in reverse.
We characterise our $\Qcir\leftrightarrow\Qaux$ mapping sequence by performing a modification of single-qubit \ac{RBM}, in which we alternate Clifford operations on the $\Qcir$ and $\Qaux$ qubits, as illustrated in \extfig{extfig:hyperfinetransfer}(c).
We assume that (i) the single-qubit gate errors for the $\Qcir$ and $\Qaux$ qubits are negligible compared to the $\Qcir\leftrightarrow\Qaux$ transfer infidelity (we typically observe single-qubit gate errors \ish\num{1e-4} for the \Ca{} hyperfine qubits), and (ii) the fidelity of the transfer $\Qcir\rightarrow\Qaux$ is similar to $\Qaux\rightarrow\Qcir$.
We therefore we model the survival probability as
\begin{equation*}
    S(m)=\frac{1}{2}+Bp^m
\end{equation*}
where $m$ is the number of hyperfine transfers, $B$ accounts for \ac{SPAM} error offsets, and $p$ is the depolarising probability for the transfer, related to the error per transfer as
\begin{equation*}
    \epsilon_{\cir\leftrightarrow\aux}=\frac{1-p}{2}.
\end{equation*}
The \ac{RBM} results are shown in \extfig{extfig:hyperfinetransfer}(c); we measure an error per transfer of \alicepertransfer{} (\bobpertransfer{}) for Alice (Bob).

\subsection{Conditional Operations}\label{methods:conditionaloperations}
To complete the \ac{QGT} protocol, the two modules perform mid-circuit measurements of the \network{} qubits, exchange the measurement outcomes, and apply a local rotation of their \circuit{} qubits conditioned on the outcomes of the measurements.
By virtue of the spectral isolation between the two species of ions, mid-circuit measurements of the \network{} qubits can be made without affecting the quantum state of the \circuit{} qubits.
The mid-circuit measurement outcomes, $m_\mathrm{A}, m_\mathrm{B}\in\{0,1\}$, are exchanged in real-time via a classical communication channel between the modules -- in our demonstration, this is a TTL link connecting the control systems of the two modules.
Following the exchange of the measurement outcomes, the modules, Alice and Bob, perform the conditional rotations $U_\mathrm{A}$ and $ U_\mathrm{B}$, respectively, where
\begin{align*}
    U_\mathrm{A} &= \begin{cases}
        S^\dagger & \text{if } m_\mathrm{A}\oplus m_\mathrm{B} = 0,\\
        S & \text{otherwise},\\
    \end{cases}\\
    U_\mathrm{B} &= \begin{cases}
        S & \text{if } m_\mathrm{A}\oplus m_\mathrm{B} = 0,\\
        S^\dagger & \text{otherwise},\\
    \end{cases}\\
\end{align*}
where $S=\mathrm{diag}(1, \mathrm{i})$.
Errors in the mid-circuit measurements of the \network{} qubits will result in the application of the wrong conditional rotation; effectively, this would appear as a joint phase-flip of the \circuit{} qubits following the teleported gate.
The mid-circuit measurement errors arise from the non-ideal single-qubit rotation of the \network{} qubit to map the measurement basis onto the computational basis, and errors due to the fluorescence detection of the \network{} qubit.
Using \ac{RBM}, we measure single-qubit gate errors for the \network{} qubits of \alicesrrotation{} and \bobsrrotation{} for Alice and Bob, respectively.
The error in the fluorescence detection is estimated from the observed photon scattering rates of $\Q{\net{}}$ states, in addition to the $\approx\SI{390}{\milli\second}$ lifetime of the $\ket{\up{\net{}}}$ state~\cite{letchumanan_lifetime_2005}.
We choose a mid-circuit measurement duration of \midcircuitmeasurementduration{}; we estimate fluorescence detection errors of \alicesrfluoerror{} and \bobsrfluoerror{} for Alice and Bob, respectively.
Combining these error mechanisms, we estimate contributions to the teleported \ac{CZ} gate error of \alicemidcircuiterror{} and \bobmidcircuiterror{} for Alice and Bob, respectively.
\begin{acronym}
    \acro{DQC}{distributed quantum computing}
    \acro{QGT}{quantum gate teleportation}
    \acro{QST}{quantum state tomography}
    \acro{QPT}{quantum process tomography}
    \acro{QCCD}{quantum charge-coupled device}
    \acro{LOCC}{local operations and classical communication}
    \acro{CZ}{controlled-Z}
    \acro{EIT}{electromagnetically-induced transparency}
    \acro{CP}{completely positive}
    \acro{CVQC}{continuous-variable quantum computing}
    \acro{IPE}{ion-photon entanglement}
    \acro{SDF}{spin-dependant force}
    \acro{RBM}{randomised benchmarking}
    \acro{SPAM}{state-preparation and measurement}
    \acro{POVM}{positive-operator-valued measure}
\end{acronym}

\bibliography{library}
\clearpage
\section{Acknowledgements}
We thank Oana B\v{a}z\v{a}van, Sebastian Saner, and Donovan Webb for maintenance of the 674-nm laser system.
We thank Chris Ballance and Laurent Stephenson for design of the apparatus, P\'{e}ter Juh\'{a}sz for comments on the manuscript, Sandia National Laboratories for supplying the ion traps used in this experiment, and the developers of the control system ARTIQ~\cite{ARTIQ}.
\dougal{} acknowledges support from the U.S.\ Army Research Office (ref.\ W911NF-18-1-0340).
\davidN{} acknowledges support from Merton College, Oxford.
\ellis{} acknowledges support from the U.K.\ EPSRC \enquote{Quantum Communications} Hub EP/T001011/1.
\raghu{} acknowledges funding from an EPSRC Fellowship EP/W028026/1 and Balliol College, Oxford.
\gabriel{} acknowledges support from Wolfson College, Oxford.
This work was supported by the U.K.\ EPSRC \enquote{Quantum Computing and Simulation} Hub EP/T001062/1.

\section{Author Information}

\subsection{Contributions}
\dougal{}, \peter{}, \davidN{}, \ellis{}, \ayush{}, \beth{}, \raghu{}, \gabriel{} built and operated the experimental apparatus.
\dougal{} led the experimental work, with assistance from \peter{} and \davidN{}.
\dougal{} performed the data analysis and prepared the manuscript with input from all authors.
\davidL{} secured funding and supervised the work.
All authors contributed to the discussion and interpretation of results.

\subsection{Competing Interests}
\raghu{} is partially employed by Oxford Ionics Ltd. The remaining authors declare no competing interests.

\subsection{Corresponding Authors}
Correspondence should be addressed to \dougal{} or \davidL{}. 
\clearpage
\setcounter{figure}{0}
\renewcommand{\figurename}{Ext.~Fig.}
\renewcommand{\tablename}{Ext.~Fig.}

\begin{figure*}[t]
    \centering
    \includegraphics[width=178mm]{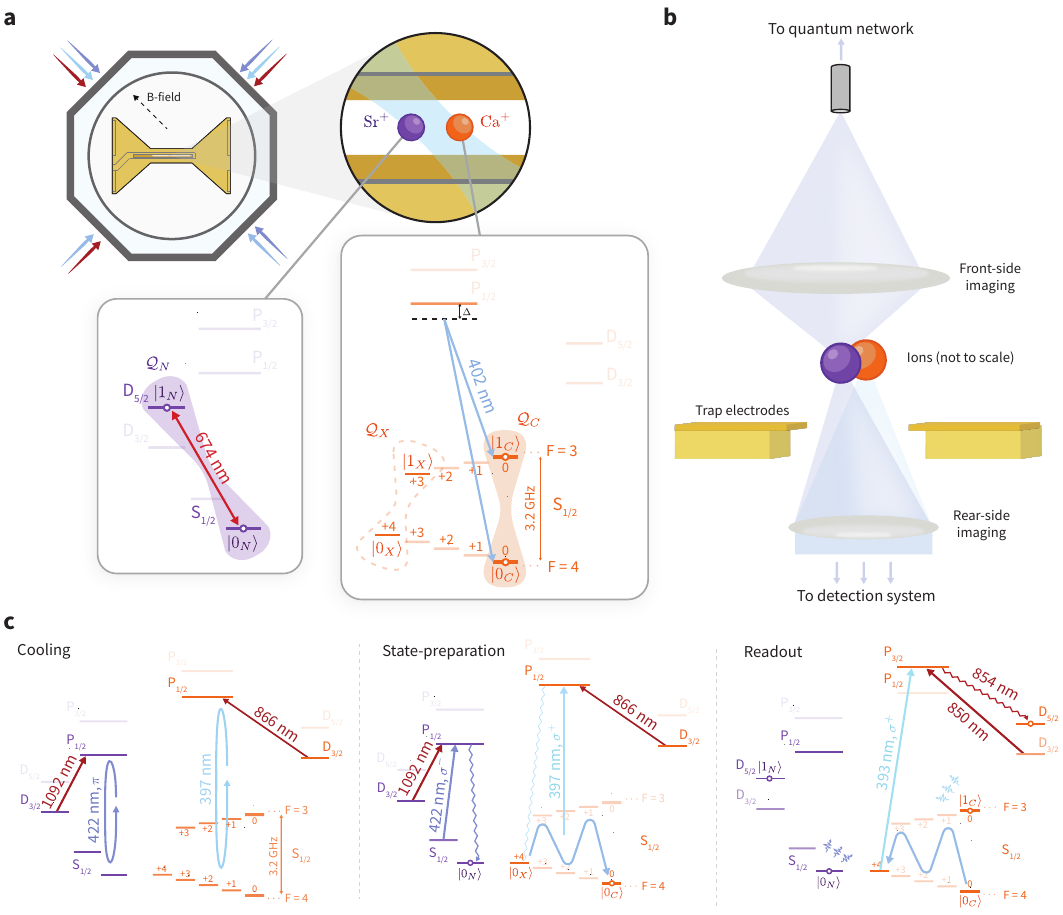}
    \caption{%
    \justifying
    \textbf{Outline of a trapped-ion module.}
    \textbf{a}, An ultra-high vacuum chamber houses a micro-fabricated surface Paul trap, which co-traps one \Sr{88} ion and one \Ca{43} ion.
    The ions are manipulated using lasers, which are delivered parallel to the surface of the trap.
    The \Sr{} ion provides an optical \network{} qubit, $\Qnet{}$, which is coherently manipulated using a \SI{674}{\nano\meter} laser.
    The ground hyperfine manifold of the \Ca{} ion provides a \circuit{} qubit, $\Qcir{}$, and an \auxiliary{} qubit, $\Qaux{}$.
    The qubits in the ground hyperfine manifold are addressed using a pair of \SI{402}{\nano\meter} Raman beams.
    \textbf{b}, The rear-side imaging system is used to perform fluorescence detection for qubit readout of both species.
    The front-side imaging system is used for single-photon collection from the \Sr{} ion during the generation of entanglement.
    A high-numerical aperture (0.6 NA) lens couples the single-photons into a single-mode optical fibre, which connects to the optical quantum network.
    Both imaging systems are outside the vacuum chamber.
    \textbf{c}, Energy level diagrams for cooling, state-preparation, and readout of each species.
    For state preparation and readout of the \circuit{} qubit $\Q{\cir{}}$ in \Ca{}, we utilise the \auxiliary{} qubit $\Q{\aux{}}$. During state preparation, we prepare $\ket{\down{\aux{}}}$ via optical pumping and then transfer it to $\ket{\down{\cir{}}}$ using Raman $\pi$-pulses. For readout, we transfer $\ket{\down{\cir{}}}$ to $\ket{\down{\aux{}}}$ before shelving to the $\D{5/2}$ manifold.
    Fluorescence detection is then used for both species, to indicate whether the ion is shelved.
    }
    \label{extfig:apparatus}
\end{figure*}

\begin{figure*}[t]
    \centering
    \begin{tblr}{
  colspec = {cccc},
  row{1} = {c},
  cell{2}{1} = {r=3}{m}, % Multi-row cell starting from the second row
  cell{5}{1} = {r=3}{m}, % Multi-row cell starting from the fifth row
}
\hline[2pt]
Module & Qubit & $\epsilon_{\ket{0}}$ ($\times 10^{-3}$) & $\epsilon_{\ket{1}}$ ($\times 10^{-3}$) \\ %& SPAM Error \\
\hline[1pt]
Alice & Network qubit, $\Qnet{}$ & \alicespamnetdown{} & \alicespamnetup{} \\ %& \alicespamnet{} \\
      & Circuit qubit, $\Qcir{}$ & \alicespamcirdown{} & \alicespamcirup{} \\ %& \alicespamcir{} \\
      & Auxiliary qubit, $\Qaux{}$ & \alicespamauxdown{} & \alicespamauxup{} \\ %& \alicespamaux{} \\
\hline[1pt]
Bob & Network qubit, $\Qnet{}$ & \bobspamnetdown{} & \bobspamnetup{} \\ %& \bobspamnet{} \\
      & Circuit qubit, $\Qcir{}$ & \bobspamcirdown{} & \bobspamcirup{} \\ %& \bobspamcir{} \\
      & Auxiliary qubit, $\Qaux{}$ & \bobspamauxdown{} & \bobspamauxup{} \\ %& \bobspamaux{} \\
\hline[2pt]
\end{tblr}
    \caption{%
    \justifying
    \Acl{SPAM} errors for all of the qubit states, in each module. The average SPAM error is \avgspam{}.}
    \label{exttab:spam}
\end{figure*}

\begin{figure*}[t]
    \centering
    \includegraphics[width=178mm]{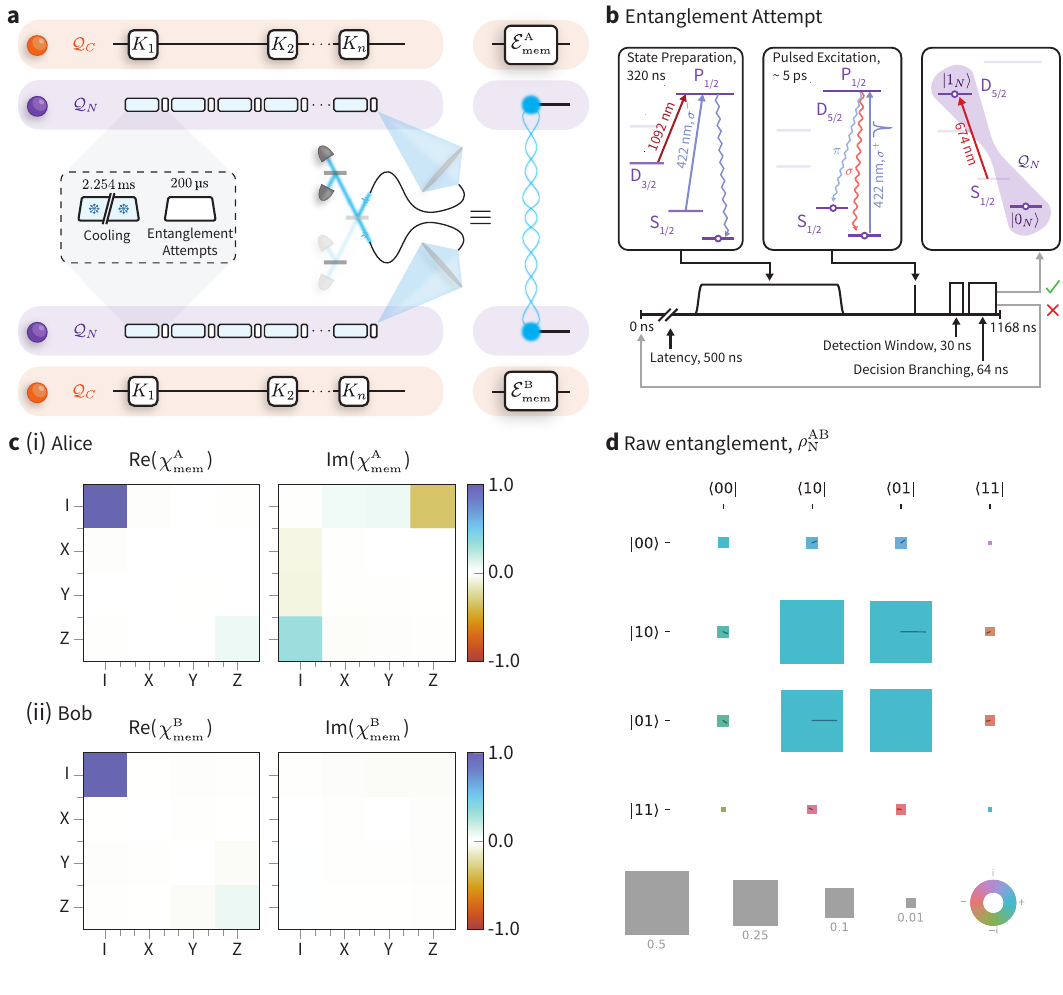}
    \caption{%
    \justifying
    \textbf{Generation of remote entanglement and robust memory of the \circuit{} qubits.}
    \textbf{a}, Entanglement is generated between the \network{} qubits using \tryduration{} of entanglement attempts interleaved with \betweeninteractions{} of sympathetic re-cooling using the \Sr{} ion.
    This is repeated until the entanglement is successfully heralded by a particular detector click pattern.
    While attempting to generate entanglement between the \network{} qubits, Knill dynamical decoupling pulses, $K_i$, are used to preserve the state of the \circuit{} qubits.
    \textbf{b}, Each entanglement attempt has a total duration of \cyclelength{}.
    We perform a $\SI{320}{\nano\second}$ state-preparation pulse (which has a switching latency of $\SI{500}{\nano\second}$), pumping the \Sr{} ion into the lower ground Zeeman state.
    A $\sim\SI{5}{\pico\second}$ pulse excites the \Sr{} ion to the upper $\P{1/2}$ level, which rapidly decays to one of the ground Zeeman levels (lifetime $\sim\SI{7}{\nano\second}$), thereby generating ion-photon entanglement.
    We collect a photon from each of the modules, interfere them on a beamsplitter, and perform a projective measurement on the two-photon polarisation state.
    Particular detector click patterns occuring within the detection window herald the successful generation of remote entanglement.
    We then exit the attempt loop and map the entanglement into the optical \network{} qubits, $\Qnet{}$.
    \textbf{c}, Reconstructed process matrices for the process of storing the state of the \circuit{} qubit in (i) Alice and (ii) Bob while generating entanglement on the \network{} qubits.
    The reconstructed process matrices have fidelities \alicememoryfidelity{} and \bobmemoryfidelity{} for Alice and Bob, respectively.
    \textbf{d}, Reconstructed density matrix of the remotely entangled \network{} qubits.
    The state indicates a fidelity of \rawfidelity{} to the $\ket{\Psi^+}$ Bell state.
    }
    \label{extfig:entanglementgeneration}
\end{figure*}

\begin{figure*}[t]
    \centering
    \includegraphics[width=178mm]{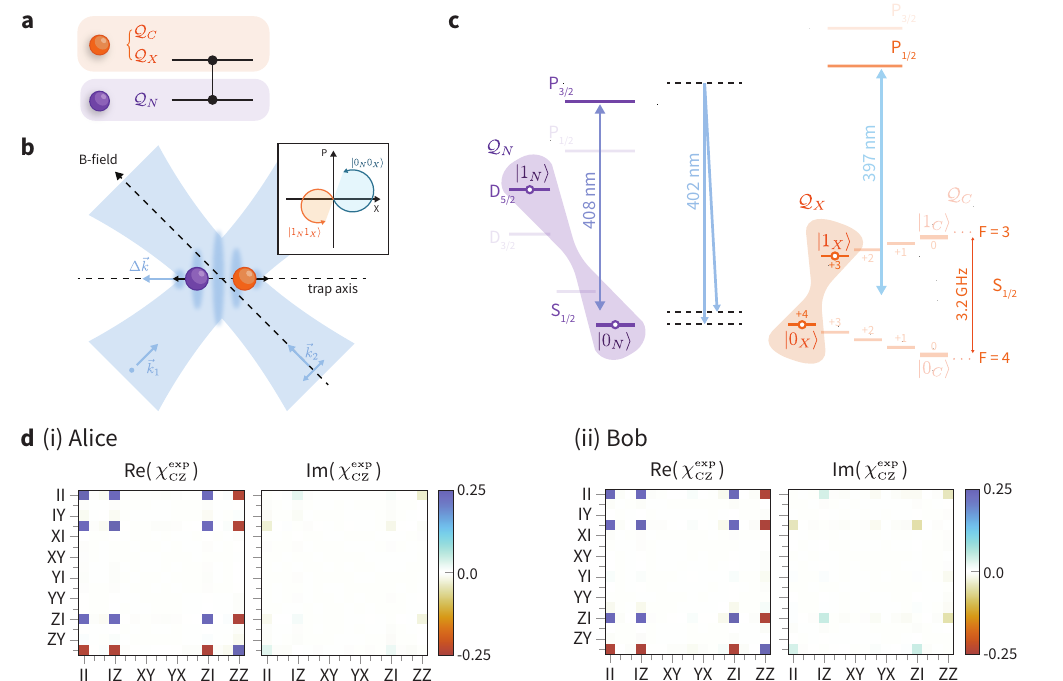}
    \caption{%
    \justifying
    \textbf{Implementation of the local mixed-species \ac{CZ} gates.} 
    \textbf{a}, Circuit element for the local \ac{CZ} gate, implemented between the $\Qnet{}$ and $\Qaux{}$ qubits. 
    \textbf{b}, Geometry for the mixed-species gate mechanism. 
    A pair of Raman lasers are aligned orthogonal to one another, such that their relative wavevector, $\Delta\vec{k}$ is along the trap axis. 
    The interference of these beams gives rise to a polarisation travelling-standing wave which induces spin-dependent light-shifts oscillating at a frequency close to the frequency of the axial out-of-phase motional mode. 
    The ions therefore experience a spin-dependent force that displaces the spin states in phase space, as depicted in the inset, thus enabling the implementation of geometric phase gates.
    \textbf{c}, Energy level diagram for the gate mechanism acting on the $\Qnet{}$ and $\Qaux{}$ qubits. By tuning the Raman lasers to \SI{402}{\nano\meter}, we couple to both the \SI{397}{\nano\meter} $\S{1/2}\leftrightarrow\P{1/2}$ dipole transition in \Ca{} and the \SI{408}{\nano\meter} $\S{1/2}\leftrightarrow\P{3/2}$ dipole transition in \Sr{}.
    \textbf{d} Process matrices for the local, mixed-species \ac{CZ} gates for (i) Alice and (ii) Bob. The process matrices have average gate fidelities of $\alicewzzfidelity{}$ and $\bobwzzfidelity{}$ for Alice and Bob, respectively.
    }
    \label{extfig:mixedspeciesgates}
\end{figure*}

\begin{figure*}[t]
    \centering
    \includegraphics[width=178mm]{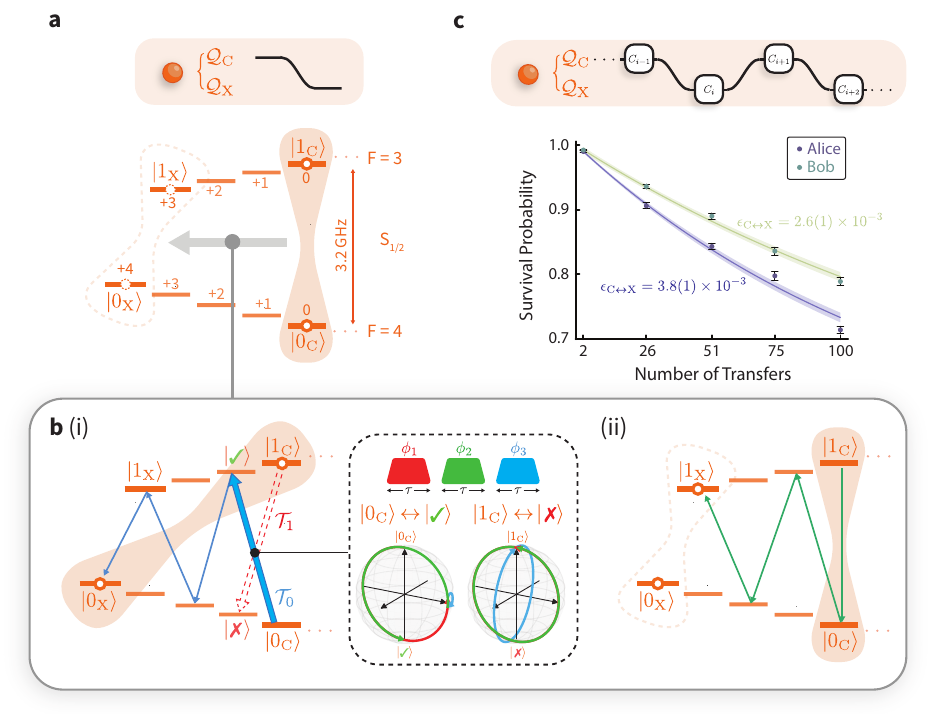}
    \caption{%
    \justifying
    \textbf{Transfer between the \circuit{} and \auxiliary{} qubits.}
    \textbf{a}, Circuit element and level diagram depicting the coherent transfer of quantum information from the $\Qcir{}$ qubit to the $\Qaux{}$ qubit. 
    The inverse transfer is implemented by performing the same steps in reverse.
    \textbf{b}, The transfer pulse sequence comprises two steps.
    (i) The first step maps the state $\ket{\down{\cir{}}}$ to $\ket{\down{\aux{}}}$. 
    Due to the near-degeneracy of the intended transition $\mathcal{T}_0:\ket{\down{\cir{}}}\leftrightarrow\ket{\text{\textcolor{green}{\ding{51}}}}$ (thick blue arrow) and the unwanted transition $\mathcal{T}_1:\ket{\up{\cir{}}}\leftrightarrow\ket{\text{\textcolor{red}{\ding{55}}}}$ (red dashed arrow), separated by only $\ish\SI{15}{\kilo\hertz}$, we employ a composite pulse sequence to suppress off-resonant coupling to the $\mathcal{T}_1$ transition. 
    The composite pulse sequence, shown in the dashed box, comprises 3 pulses of duration $\tau$ resonant with the $\mathcal{T}_0$ transition with differing phases $\phi_i$. The pulse duration, $\tau$, is equal to the $2\pi$-time of the $\mathcal{T}_1$ transition, $\phi_1=\phi_3=0$, and $\phi_2\approx 2\pi\times0.231$ is optimised experimentally.
    The subsequent transfer pulses (thin blue arrows) are $\pi$-pulses on the relevant transitions.
    This sequence therefore performs the mapping $\ket{\down{\cir{}}}\rightarrow\ket{\down{\aux{}}}$, while leaving the state $\ket{\up{\cir{}}}$ unaffected.
    (ii) The second step comprises a sequence of $\pi$-pulses which maps $\ket{\up{\cir{}}}\rightarrow\ket{\up{\aux{}}}$.
    This completes the coherent transfer $\Qcir{}\rightarrow\Qaux{}$.
    \textbf{c}, The performance of the transfer sequence is characterised using a modified version of \ac{RBM} in which we alternately perform Clifford operations on the $\Qcir$ and $\Qaux$ qubits.
    By measuring the survival probability for different numbers of transfers, and neglecting the errors of the single-qubit gates $C_i$ (which are $\sim 1\times 10^{-4}$), we extract the error per transfer, $\epsilon_{\mathrm{C}\leftrightarrow\mathrm{X}}$, yielding \alicepertransfer{} and \bobpertransfer{} for Alice and Bob, respectively.
    }
    \label{extfig:hyperfinetransfer}
\end{figure*}

\end{document}